\documentclass[11pt]{article}
\usepackage{amssymb}

\usepackage{amsmath}
\usepackage{amsbsy}

\input{tcilatex}

\begin{document}

\begin{center}
\textbf{Geometrization of Classical Physics, Unified for Electrodynamics and
Gravitation}

Jos\'{e} G. Vargas\\[0pt]
October 2018
\end{center}

\vspace*{.5cm}

\hfill \textit{To the memory of the late Dr. Marsha Torr}

\vspace*{1.5cm}

\textbf{Abstract.} We show the most unexpected result that the form of the
Lorentz force is inescapable in Finsler geometry, as it is always present in
the autoparallels, even in particular in the natural liftings to the Finsler
bundle of arbitrary connections with torsion in the standard bundle of
spacetime (Although these liftings retain the form $R_{\,\nu \lambda }^{\mu
}\omega ^{\nu }\wedge \omega ^{\lambda }$ of the lifting, the soldering
forms, $\omega ^{\mu }$, have changed, and so do correspondingly change the
components $R_{\,\nu \lambda }^{\mu }).$ The $q/m$ factor is to be viewed as
an expected effect of the asymmetry inherent in the ``particle in a field
picture''.

Finslerian torsions, $\Omega ^{\mu }=d\omega ^{\mu }-\omega ^{\nu }\wedge
\omega _{\nu }^{\mu }=R_{\,\nu \lambda }^{\mu }\omega ^{\nu }\wedge \omega
^{\lambda }+S_{\,\nu i}^{\mu }\omega ^{\nu }\wedge \omega _{0}^{i}$, $%
(\lambda =(0,l_{i})=0,1,2,3)$, span three sectors: (a) electrodynamic,
defined by $\Omega ^{i}=0$ and $S_{\nu l}^{0}=0$, (b) ``dark matter'' (for
lack of a better name), defined by $\Omega ^{i}=0,$ $S_{\nu l}^{i}\neq 0$
(It affects the equation of the autoparallels with additional terms, not
only for the force but also for the momentum) and (c) the dark sector of the 
$\Omega ^{i}$, ``dark'' because it contributes to the energy equations but
not to the equations of the motion.

We then assume teleparallelism. The linearization of the first Bianchi
identity for torsions of type (a) becomes the first pair of Maxwell's
equations. And the vanishing of a curvature related vector-valued
differential 3-form, and the splitting of the connection into contorsion and
Levi-Civita differential 1-form yields Einstein's equations as a statement
of annulment of a sum of three currents, metric dependent (gravitational),
torsional (other fields and matter) and a third one that mixes torsion and
metric and with cosmological flavor.

\section{Introduction}

In 1905, Einstein did the best that was possible at the time to deal with
the issue of finding the dynamics underlying electromagnetic theory, which
certainty was not classical mechanics. Present mathematics allows us to do
much better, thanks to more encompassing mathematics, like Finsler bundles
in this paper. A by-product of these bundles is that the unification of
gravity with electromagnetism is not to be sought but actually unavoidable.

Finsler bundles can be seen as refibrations of standard bundles. If, in
addition to refibrating the spacetime's standard frame bundle, one lifts the
connection to the new bundle, the form of the Lorentz force emerges in the
equations of the autoparallels for components of type $R$ of $\Omega ^{0}$.
This feature that emerges in the refibration was not possible in
pre-Finslerian differential geometry, since $\Omega ^{0}$ does not remain
unchanged under Lorentz transformations. This does not necessarily clash
with the Lorentz group. It simply happens that the boosts without rotation
act on ``reduced vectors'' (four dimensional) tangent to the 7-D base space
of the Finslerian bundle of spacetime. Putting things differentially, the
group in the fibers of this bundle is the rotation group in three
dimensions. $\Omega ^{0}$ thus behaves as a scalar field in 7-D under
tangent Lorentz transformations, the boosts simply move the scalar-valued
differential form $\Omega ^{0}$ from one point to another in that 7-D
manifold. Although the effect of the metric certainly does not vanish, it
recedes to the background for this special torsion in this (Finslerian)
setting. furthermore, it has not been realized -- mostly because one has not
cared -- that Finsler bundles are canonical for the Lorentz signature, like
the Riemannian bundles are canonical for the positive-definite signature.
This is so because of the following. Spacetime tangent vectors are lifted to
the Finsler bundle as equivalence classes of 7-D tangent vectors. The
equivalence puts the classes in a one-to-one correspondence with 4-vectors.
This takes place by first choosing a special dimension to define those
equivalence classes. But there are not special directions when the metric is
positive definite. Temporal dimensions are special. Not counting
null-dimensions, there is one and only one special direction in any set of
four independent spacetime directions.

We shall have done in this paper enough to start giving credence to
Einstein's thesis of logical homogeneity of differential geometry and
theoretical physics \cite{R38}. From the unheeded recommendations that E.
Cartan gave to Einstein on the issue of geometrization with teleparallelism,
this thesis should be seen as implying that the field equations and other
key physical equations are the equations of structure and Bianchi identities
of some differential geometry that he did not specify. It has not been
noticed that we live at a moment of history where the mathematics can guide
us beyond our wildest expectations, once we have assumed that the connection
of spacetime is Finslerian, rather than pseudo-Riemannian or
pseudo-Riemannian with torsion. We said ``guide us'', since differential
geometry is not the last word. But we can bring its equations closer to the
basic equations of the K\"{a}hler version of quantum mechanics (Since his
papers are in German, we ask impatient readers to google ``Alterman 2016'',
``Alterman 2017'' and ``Alterman 2018'' to catch a glimpse of that version).
By this author's next paper, we shall have started to use in classical
physics the calculus that he used for quantum mechanics (KC\ for K\"{a}hler
calculus). In it, we shall view electrodynamics as a theory of the
electromagnetic field created by charges, but where both field and
charge-current are manifestations of the torsion of space-time.

The exploits of the paradigm and the extraordinary precision of
electrodynamics would seem to constitute a strong argument against an effort
such as the present one (Let us simply retort that the Newtonian solar
system also was very precise). But there are very important basic problems
in the paradigm, which are simply ignored when not overlooked. An even worse
case is the purported justification of the absence of a gravitational
energy-momentum tensor in Einstein's theory of general relativity. In this
last case, the problem lies in that a wrong decision was made by the physics
community in the years 1917-1920, as we shall show in this paper.

We shall start by showing that the Finslerian setting allows one to
integrate the two indices of the electromagnetic field into the three
indices of the torsion. There is then the extraordinary revelation reported
in the abstract on the relation of the autoparallels (i.e. lines of constant
direction) to the Lorentz force and to the splitting of the torsion into
three sectors of interest. The use of Finsler bundles is not just a matter
of trusting the potential for physics of the results that we shall obtain in
this and coming papers. We shall then use teleparallelism (TP) in those
bundles to deal with field equations, Einstein's and first pair of
Maxwell's. The second pair requires replacing the exterior calculus with the
KC in differential geometry. Differential forms, not tensor fields,
constitute the words of the language to deal with sophisticated differential
geometry and with the KC. In particular, the replacement of the concept of
energy-momentum tensors with the concept of vector-valued differential
3-form reveals subtleties that the opaque tensor calculus cannot, specially
in connection with the issue of gravitational energy-momentum tensors.

Except for global issues, the TP requirement is equivalent to having the
curvature equal to zero (We are using the term global in the sense of the
mathematicians, meaning that something is not global if it fails to apply at
every point of a differentiable manifold, even if it fails at just a single
point). We shall use the term curvature to refer to $d\omega _{\mu }^{\nu
}-\omega _{\mu }^{\lambda }\wedge \omega _{\lambda }^{\nu }$, which is not
the Riemannian curvature unless the connection is Levi-Civita's$.$ We shall
follow Cartan and use for $d\omega _{\mu }^{\nu }-\omega _{\mu }^{\lambda
}\wedge \omega _{\lambda }^{\nu }$ the term Euclidean rather than affine
curvature, when there is a metric, which restricts the affine bundle to a
Euclidean (or pseudo-Euclidean) one. Both this curvature and the Riemannian
curvature can be simultaneously present in a differentiable manifold, each
in its respective role(s). When Einstein formulated general relativity in
1915, this theory concerned the Riemannian curvature, since the concepts of
affine, Euclidean and Lorentzian curvatures were then not even known.

Einstein gets the credit for being the first to try TP for physics. He did
so in the late 1920's \cite{R39}. He failed in his attempt because the state
of the geometric art was not sufficiently developed for the task, and
because he did not listen to ---perhaps he did not understand--- E. Cartan
when they extensively corresponded on TP. He thus did not realize the
potential of TP for physics, which Cartan realized even though one could not
yet implement at the pre-Finslerian level.

These issues are best understood in the language of differential forms.
Because of its absence in the mathematical and physical literature, we
discuss in detail the circumstances that lead to the unfortunate adoption of
the Levi-Civita (LC) connection by the original theory of general
relativity. Born in 1917, the LC connection was obviously absent in the
Riemannian geometry of 1915. Assumptions were made a the beginning of the
twentieth century which would not have been made so readily if one had known
better mathematics. That is the main point that his paper makes, at the same
time as it shows the way to an obvious alternative.

\section{A bird's eye view of Finslerian differential geometry}

In 1922, Cartan stated: ``The metric does not contain all the geometric
structure of the space ... one may define the space by its structure
equations'' \cite{R10}. These equations constitute a differential system.
When integrable, the connection equations result from its integration. We
may thus reformulate Cartan and state that the geometric reality of a space
resides in its connection, term here used to refer to $(\omega ^{\mu
},\omega _{\nu }^{\lambda })$ and not just to the $\omega _{\nu }^{\lambda }$%
. Hence the view of Finsler geometry from a metric perspective is adequate
for global issues, but is totally insufficient for general purposes
(Practitioners of global differential geometry may nevertheless use the
canonical connection of the metric for their purposes).

Differential forms constitute the appropriate tool for differential
geometry. To our knowledge, only American practitioners of Finsler geometry
use differential forms, but they are only interested in global problems.
That leaves only the Cartan-Clifton formulation of Finsler geometry for our
purposes. That can be found in \cite{R73}, where we explain that and why the
true author of the paper is the differential topologist Y. H. Clifton, and
not its nominal ones.

The basis for the Clifton approach to Finsler geometry is the seminal Cartan
paper of 1934 on the subject \cite{R19}. With great effort (as he confessed
to the present author), Clifton was able to interpret it in terms of frame
bundles. To be specific, let us consider spacetime. Its standard
pseudo-orthonormal frame bundle is a 10-D manifold fibrated over a 4-D base,
i.e. over spacetime. The corresponding Finsler bundle is the same
topological space fibrated over a 7-D manifold. The latter is the so called
sphere bundle or bundle of directions. These names are not very fortunate
because they evoke a Euclidean rather than a Lorentzian connotation. Suffice
to say that a natural set of coordinates are the $(t,x^{i},u^{i})$, where
the $u^{i}$'s are the so called velocity coordinates, or parameters of the
Lorentz boosts without rotation. On curves that are natural liftings of
spacetime curves (i.e. that satisfy the conditions $dx^{i}-u^{i}dt=0$) and $%
u^{i}$ becomes $dx^{i}/dt$. Hence the term velocity coordinates follows.

From a frame bundle perspective, the equations of structure on standard
bundles 
\begin{align}
& d\omega ^{\mu }-\omega ^{\nu }\wedge \omega _{\nu }^{\mu }=R_{\pi \rho
}^{\mu }\omega ^{\pi }\wedge \omega ^{\rho }, \\
& d\omega _{\mu }^{\nu }-\omega _{\mu }^{\lambda }\wedge \omega _{\lambda
}^{\nu }=R_{\mu \pi \rho }^{\nu }\omega ^{\pi }\wedge \omega ^{\rho }.
\end{align}%
state that the non-horizontal parts of $d\omega ^{\mu }$ and $d\omega _{\mu
}^{\nu }$ cancel out with the non-horizontal parts of $\omega ^{\nu }\wedge
\omega _{\nu }^{\mu }$ and $\omega _{\mu }^{\lambda }\wedge \omega _{\lambda
}^{\nu }$ respectively. The summations on the right hand side are not over
all $\pi $ and all $\rho $, but only over a basis of differential 2-forms $%
\omega ^{\pi }\wedge \omega ^{\rho }$. In the presence of a metric,
equations (1)-(2)have to be complemented with $\omega _{\mu \nu }+\omega
_{\nu \mu }=0$, equivalently with $\omega _{0}^{i}=\omega _{i}^{0}$ and $%
\omega _{i}^{j}=-\omega _{j}^{i}$.

The equations of structure for a general connection on the Finsler bundle of
a manifold endowed with a metric read 
\begin{align}
& d\omega ^{\mu }-\omega ^{\nu }\wedge \omega _{\nu }^{\mu }=R_{\pi
\,p}^{\mu }\omega ^{\pi }\wedge \omega ^{\rho }+S_{\pi \,i}^{\mu }\omega
^{\pi }\wedge \omega _{0}^{i}, \\
& d\omega _{\mu }^{\nu }-\omega _{\mu }^{\lambda }\wedge \omega _{\lambda
}^{\nu }=R_{\mu \,\pi \rho }^{\text{ \ }\nu }\omega ^{\pi }\wedge \omega
^{\rho }+S_{\mu \,\lambda i}^{\text{ \ }\nu }\omega ^{\lambda }\wedge \omega
_{0}^{i}+T_{\mu \,ij}^{\text{ \ }\nu }\omega _{0}^{i}\wedge \omega _{0}^{j}.
\end{align}%
The summation over the $T$ terms is for a complete set of independent $%
\omega _{0}^{i}\wedge \omega _{0}^{j}$ forms. Terms containing at least one
factor $\omega _{l}^{m}$ are absent on the right hand side. This is to say
that the left hand sides are horizontal in spite of the implicit presence of 
$\omega $'s with two Latin indices. In the case of the properly Euclidean
signature, the role of the index zero is taken by any arbitrary index. In
the Lorentzian case, the natural choice for the ``special role coordinate''
is time. This is what we mean when we state that the Lorentzian signature is
the canonical signature of Finslerian geometry.

Some remarks of interest (if not already obvious by now) are:

(a) The $\omega _{0}^{i}$'s are horizontal on sections of the Finsler
bundle, and not as linear combinations of the $\omega ^{\mu }$'s.

(b) The $\boldsymbol{e}_{\mu }$'s dual to the $\omega ^{\mu }$'s on sections
of the same bundle, as the $\omega ^{\mu }$'s themselves, do not coincide
with the $\boldsymbol{e}_{\mu }$'s and $\omega ^{\mu }$'s on sections of the
standard bundle. In particular, the four velocity, $u^{\mu }\boldsymbol{e}%
_{\mu }$, is simply $\boldsymbol{e}_{0}$ on sections of the Finsler bundle.

(c) On these sections, the $\omega ^{\mu }$'soldering forms can be written
as 
\begin{equation}
\omega ^{0}=Ldt+\Lambda _{r}\sigma ^{r},\;\ \ \ \ \ \ \ \ \ \ \;\omega
^{i}=A_{\text{ }r}^{i}\sigma ^{r},
\end{equation}%
where $\sigma ^{r}=dx^{r}-u^{r}dt$ and where $L,\Lambda _{r}$ and $A_{r}^{i}$
depend on the $x$'s and $u$'s. The second equations (5) are special in that
they are linear combinations of just the three differential forms $\sigma
^{i}.$ The equations $\sigma ^{i}=0$ are the so called natural lifting
conditions. Without further ado, let us remark that, on curves, all
differentials are linear functions of just one, say $dt$. In this case, $%
\sigma ^{i}=0$ yields $u^{i}=dx^{i}/dt.$

(d) A consequence of (b): although the $R_{\mu \lambda }^{\mu }\omega ^{\nu
}\wedge \omega ^{\lambda }$ retain this form when lifted to sections of the
Finsler bundle, this is a different set of $R_{\nu \lambda }^{\mu }$'s since
he soldering forms on sections are changed by the lifting. Assume that we
make all terms in (3)-(4) equal to zero. This would look as Riemann with
torsion geometry. It would not be so, the reason being that the $\omega ^{i}$
do not take in the usual bundle the form $A_{r}^{i}\sigma ^{r}$. These
remarks are of great importance for understanding that the form of the
Lorentz force is always present on sections of the Finsler bundle, even for
liftings of arbitrary connections on the usual bundle.

(e) On natural liftings, $\omega ^{0}=Ldt$, $\omega ^{i}=0.$ Hence, on such
curves, the metric can be replaced with 
\begin{equation}
ds=Ldt\;\ \ \ \ \ \ \ \ \ \ \ \ \ \ \ \;\text{or}\;\ \ \ \ \ \ \ \ \ \ \ \ \
\ \ \;\omega ^{0},\;\mbox{mod}\sigma ^{i}.
\end{equation}

\section{The Finslerian form of the equations of motion of general relativity%
}

In this section, we acquaint readers develop familiarity with computations
of the equations of motion in Finsler geometry, where the equations of the
autoparallels or lines of constant direction are given by 
\begin{equation}
\omega _{0}^{m}=0,\;\ \ \ \ \ \ \ \ \;\omega ^{l}=0=\sigma ^{l}.
\end{equation}%
When the torsion is zero, we shall write $\omega _{0}^{m}$ as $\alpha
_{0}^{m}$, and shall thus have 
\begin{equation}
0=d\omega ^{0}-\omega ^{i}\wedge \alpha _{i}^{0}.
\end{equation}%
We need to compute $\alpha _{i}^{0}.$Using equations (5) in (8), we obtain 
\begin{equation}
0=(d\Lambda _{i}-L_{,i}dt+\alpha _{r}^{\text{ }0}A_{\text{ }i}^{r})\wedge
\sigma ^{i}+(\Lambda _{m}-L_{.\,m})dt\wedge du^{m},
\end{equation}%
where subscripts ``,'' and ``.'' mean partial derivatives with respect to $x$
and $u$ respectively. We have used that $L_{,m}dx^{m}\wedge dt=L_{,m}\sigma
^{m}\wedge dt$. Equation (9) can further be written as 
\begin{equation}
\alpha _{i}\wedge \sigma ^{i}+(\Lambda _{m}-L_{.\,m})dt\wedge du^{m}=0,
\end{equation}%
where $\alpha _{i}$ is defined as 
\begin{equation}
\alpha _{i}\equiv d\Lambda _{i}-L_{,i}dt+\alpha _{r}^{0}A_{i}^{r}.
\end{equation}%
Since there are no $dx^{l}$ factors in $dt\wedge du^{m}$, the two terms in
(10) must be zero. From the second term, we get 
\begin{equation}
\Lambda _{m}=L_{.\,m},
\end{equation}%
The annulment of the first term implies, 
\begin{equation}
\alpha _{i}=C_{im}\sigma ^{m},\;\;\;\text{with}\;C_{im}=C_{mi}.
\end{equation}%
We now assume the equation satisfied by all natural lifting conditions,
namely $\sigma ^{m}=0.$ Thus, $\alpha _{i}=0$ and, therefore, 
\begin{equation}
0=d\Lambda _{i}-L_{,i}dt+\alpha _{r}^{0}A_{i}^{r}=dL_{.\,i}-L_{,i}dt+\alpha
_{r}^{0}A_{i}^{r},
\end{equation}%
where we have used (12). Since the autoparallels satisfy $\alpha _{0}^{r}$
when the torsion is zero, we finally have 
\begin{equation}
L_{,i}dt=dL_{.i},
\end{equation}%
or, equivalently 
\begin{equation}
\frac{\partial L}{\partial x^{i}}=\frac{d}{dt}\frac{\partial L}{\partial
u^{i}},
\end{equation}%
which are the Euler-Lagrange equations of motion in general relativity
[Recall (6)]. We have followed this route to get to Eq. (16) in order to
facilitate the understanding of the contents of the next section.

One should be aware of the fact that Eqs. (5) give the soldering form for
affine connections in Finsler geometry. We have not yet required that $%
(\omega ^{0})^{2}-\sum (\omega ^{i})^{2}$ be an invariant. Much less has one
required that this invariant be $dt^{2}-\sum_{i}(dx^{i})^{2}.$\ In the last
case, the coefficient $L$ is readily computed. We then have 
\begin{equation}
\omega ^{0}=\gamma (dt-u_{i}dx^{i})=\gamma dt-\gamma u_{i}(\sigma
^{i}+u^{i}dt),
\end{equation}%
where $\gamma =(1-u^{2})^{-1}$, and, therefore, 
\begin{equation}
L=\gamma -\gamma u_{i}u^{i}=\gamma (1-u^{2})=\sqrt{1-u^{2}}=\gamma ^{-1}.
\end{equation}%
Again in flat spacetime, we get the following standard result for the
spatial components of the kinetic 4-momentum: 
\begin{equation}
\frac{\partial L}{\partial u^{i}}=\frac{-u_{i}}{\sqrt{1-u^{2}}}=-\gamma
u_{i}=p_{i}.
\end{equation}%
We shall later make use of this.

\section{Equation of motion geometrically unified for electrodynamics and
gravitation}

We shall now consider the effect of torsion on the equations of the
autoparallels in Finsler bundles. The temporal component of the torsion is
to be read from (3). The $\omega ^{\mu }$'s constitute the basis of
differential 1-forms dual to the basis of tangent vectors that one may be
using.

In order to remove clutter in the process of identifying the presence of the
Lorentz force, we assume $S_{\mu i}^{0}$ equal to zero. We also redefine the
components of the torsion as implied by the equation%
\begin{equation}
\Omega ^{0}=R_{\text{ }\mu \nu }^{0}(\omega ^{\mu }\wedge \omega ^{\nu })=R_{%
\text{ \ }0i}^{\prime \,0}dt\wedge dx^{i}+\frac{1}{2}R_{\text{ \ }%
lm}^{\prime \,0}dx^{l}\wedge dx^{m}.
\end{equation}%
The parenthesis around $\omega ^{\mu }\wedge \omega ^{\nu }$ means that we
are summing over a basis of independent differential 2-forms of this type.
On the right hand side, on the other hand, we are summing over all and $m$.
We shall later used $\omega _{0}^{i}=\omega _{i}^{0}$ and $\omega
_{i}^{j}=-\omega _{j}^{i}$, since we have not done anything to change the
basis of tangent vectors.

At this point, we have, instead of equation (8), 
\begin{equation}
0=d\omega ^{0}-\omega ^{i}\wedge \omega _{i}^{0}-R_{\text{ \ }0i}^{\prime
0}dt\wedge dx^{i}-\frac{1}{2}R_{\text{ \ }lm}^{\prime 0}dx^{l}\wedge dx^{m}.
\end{equation}%
The development of the first two terms in (21) gives rise to the right hand
side of (9), with $\alpha _{r}^{0}$ replaced with $\omega _{r}^{0}$. We now
proceed to develop the $dx$'s in terms of the $\sigma $'s, recalling that $%
dx^{i}=\sigma ^{i}+u^{i}dt.$

We write the third term on the right hand side of (21) as $-R_{\text{ \ }%
0i}^{\prime \,0}dt\wedge \sigma ^{i}$. The fourth term evolves as follows: 
\begin{align}
& -\frac{1}{2}R_{\text{ \ }lm}^{\prime \,0}(\sigma ^{l}+u^{l}dt)\wedge
(\sigma ^{m}+u^{m}dt)  \notag \\
& \qquad =-\frac{1}{2}[R_{\text{ \ }lm}^{\prime \,0}\sigma ^{l}\wedge \sigma
^{m}-R_{\text{ \ }lm}^{\prime \,0}\sigma ^{l}\wedge u^{m}dt-R_{\text{ \ }%
lm}^{\prime \,0}u^{l}dt\wedge \sigma ^{m}].
\end{align}%
On the right of (22), let us collect all the terms where $\sigma ^{1}$ is a
factor$.$ We get $(R_{\text{ \ }12}^{\prime \,0}-R_{\text{ \ }31}^{\prime
\,0}\sigma ^{3})\wedge \sigma ^{1}$. Doing the same with $\sigma ^{2}$ and $%
\sigma ^{3}$, we shall have counted each term twice. Hence the sum of all
the $\sigma ^{l}\wedge \sigma ^{m}$ terms yields 
\begin{equation}
\frac{1}{2}(R_{\text{ \ }ij}^{\prime \,0}\sigma ^{j}-R_{\text{ }ki}^{\prime
\,0}\sigma ^{k})\wedge \sigma ^{i},
\end{equation}%
with sum over cyclic permutations. In the remainder of the right hand side
of (22), the factor $\sigma ^{1}$ will emerge in 
\begin{equation*}
(R_{\text{ \ }12}^{\prime \,0}u^{2}-R_{\text{ \ }31}^{\prime
\,0}u^{3})dt\wedge \sigma ^{1}.
\end{equation*}%
Hence, when the contributions of $\sigma ^{2}$ and $\sigma ^{3}$ are also
taken into account, we get 
\begin{equation}
(R_{\text{ \ }ij}^{\prime \,0}u^{j}-R_{\text{ \ }ki}^{\prime
\,0}u^{k})dt\wedge \sigma ^{i}.
\end{equation}%
By bringing all these results to (21), we obtain 
\begin{align}
0& =[d\Lambda _{i}-L_{,i}dt+\omega _{r}^{0}A_{i}^{r}-R_{\text{ \ }%
0i}^{\prime \,0}dt+(R_{\text{ \ }ij}^{\prime \,0}u^{j}-R_{\text{ \ }%
ki}^{\prime \,0}u^{k})dt  \notag \\
& +\frac{1}{2}(R_{\text{ \ }ij}^{\prime \,0}\sigma ^{j}-R_{\text{ \ }%
ki}^{\prime \,0}\sigma ^{k})dt]\wedge \sigma ^{i}+(\Lambda
_{m}-L_{.\,m})dt\wedge du^{m}.
\end{align}%
Now as before this implies that $\Lambda _{m}=L_{.\,m}$ and that the square
bracket must be a linear combination of the $\sigma ^{m}$'s, i.e. $%
c_{mi}\sigma ^{m}$ with $c_{mi}=c_{im}$. Since the autoparallels satisfy the
conditions $\sigma ^{l}=0$ and $\omega _{0}^{r}=0,$ with $\omega
_{0}^{r}=\omega _{r}^{0}$, we conclude that they satisfy the equation 
\begin{equation}
0=dL_{.\,i}-[L_{,i}+R_{\text{ \ }0i}^{\prime \,0}+(R_{\text{ \ }ki}^{\prime
\,0}u^{k}-R_{\text{ \ }ij}^{\prime \,0}u^{j})]dt.
\end{equation}%
This is the equation of motion unified for the gravitational and
electromagnetic forces if we identity $R_{\text{ \ }0i}^{\prime \,0}$ with $%
(q/m)E)_{i}$, and $R_{\text{ \ }ij}^{\prime \,0}$ and $R_{\text{ \ }%
ki}^{\prime \,0}$ with $(q/m)B_{k}$ and $(q/m)B_{j}$, respectively.

Some readers may be thinking that the factor $q/m$, which differs from
particle to particle, will invalidate the argument that the $R$ part of the
torsion is to be identified with the electromagnetic field. The right
interpretation is that the ``outside of the particle'' electromagnetic field
must indeed be identified with the torsion, since, at the position of the
particle, the torsion is not dictated by the outside field but will be the
torsion that defines the specific particle. It is a challenge for this
emerging theory the obtaining of the factor $q/m$ from the torsion field.
But the right way would be to do so after contact has been made with the
quantum mechanics consistent with this geometric picture for classical
physics. We thus read from this result that%
\begin{equation}
R_{-\mu \nu }^{0}=F_{\mu \nu }=\left[ 
\begin{array}{cccc}
0 & E_{x} & E_{y} & E_{z} \\ 
-E_{x} & 0 & -B_{z} & B_{y} \\ 
-E_{y} & B_{z} & 0 & -B_{x} \\ 
-E_{z} & -B_{y} & B_{x} & 0%
\end{array}%
\right]
\end{equation}%
without a $q/m$ factor.

It is understood that the basis of differential forms for the $R$'s and the $%
F$'s will be the same, which in our computation became the ($dt,$ $dx^{i}$),
as per equation (20). For this reason, we no longer need to keep the primes.
These relations are basis independent, except that $\Omega ^{0}$ pertains to
canonical bases, i.e. orthonormal when both a metric and a connection define
a structure.

As for bases of differential to be used, we make all the simplifications
possible in order to establish the relation between geometric and physical
formulas. The present problem consist in identifying geometric equations
with the field equations of electrodynamics, making abstraction for the
moment of the presence of mass, charge and current. As we obtain further
corroboration of (27), we become increasingly independent of present
formulas in the paradigm and let the mathematics speak. This course of
action will show us that terms like interactions, potentials, sectors,
cosmological \textit{constant}, \textit{etc. }are not the most suitable
terms to represent what is going on. But we shall temporarily use those
terms for enhancing the parallelism between the paradigm and the physical
theory that emerges here.

Decades ago, Ringermacher \cite{R64} and the present author \cite{R72}.
guessed a torsion which did not make sense in Riemannian with torsion
geometry. It represented a very crude version of the results obtained, and
made sense only as indicating the direction of future research. Somebody
well versed on the Finsler bundle would readily realize that our result
becomes rigorously meaningful in this bundle. That is what a differential
topologist by the name of Y. H. Clifton understood [See report in \cite{R73}%
]. It was not immediately realized by any of us that, as we have just shown,
the form of the Lorentz force is canonical in the Finsler bundle of
spacetime, which brings about the consideration that follows. In the usual
bundle, one has to find a specific torsion that could be identified ---even
if not quite correctly--- with the electromagnetic field. But all
``Riemannian connections with torsion'' can be lifted to the Finsler bundle,
where all of them yield the form of the Lorentz force in the autoparallels.
So, the lifting achieves something very special, namely that they all have
in common autoparallels of the same form, involving terms with the same form
as the Lorentz force (The difference between liftings of difference
connections lies in details of the terms within the force, not in its form;
recall that the lifting of a standard connection to the Finsler bundle
changes the coefficients $R_{\nu \lambda }^{\mu }$)

Because the electromagnetic field is confined to the temporal component of
the tangent index of the torsion, Maxwell's equations do not depend on the
connection of spacetime. Thus scalar-valued differential forms suffice for
this purpose. This confirms a remark by Cartan in 1924 \cite{R38} (as if his
remarks needed confirmation!) that the most fundamental form of Maxwell's
equations is the one in terms of integrals, not in point form. This is
consistent with the formulation of electrodynamics in terms of differential
forms, when these are viewed as functions of hypersurfaces rather than as
antisymmetric multilinear functions of vectors. Cartan went on to say that
Maxwell's equations do not contain all of classical electrodynamics. They
have to be complemented with relations involving energy and momentum, and
this is where the non-scalar-valuedness explicitly enters electrodynamics
(We said explicitly because, as we said, the vector valuedness is implicit
in $\Omega ^{0}$, which was central to this paper, even though it looked as
if only scalar-valued differential forms were needed for our dealings.)

The non-vanishing of $S_{\text{ \ }\mu i}^{\prime \,0}$ and $\Omega ^{i}$
will certainly reveal a deeper reality (See next section). This by itself
indicates that one should further explore the consequences of Finslerian
geometry with torsion, and continue the deep synergy not just between
classical electrodynamics and Finsler geometry, but also of the latter with
little known sectors of the paradigm. In the next section, we shall deal
with one of them.

\section{Equations of motion beyond Electrodynamics}

In this section, we start to obtain geometric terms beyond those of
classical field theory. They concern non-electromagnetic components of the
torsion, and one of them actually involves the acceleration, like the time
derivative of the momentum does. In order to deal with this situation, the
term \textit{sector} helps. So, we shall speak of dark matter, dark energy,
Higgs and cosmological sectors. The justification for the use of the term
Higgs sector has to do with the fact that it deals with the emergence of
mass from fields. It is of the essence already in classical physics, so that
one does not have a classical theory of particles (classical mechanics) on
the one hand, and a classical theory of fields on the other. The use of the
term cosmological sector is due to the fact that the cosmological term seem
to be just the tip of a cosmological iceberg. In this section, we have said
all what we have to say about dark matter at this point. We reserve for the
next paper incipient considerations about the other three sectors.

It is not strange that the $\Omega ^{i}$ do not contribute to the equation
of the autoparallels, since $\boldsymbol{\Omega}=\Omega ^{\mu }\boldsymbol{e}%
_{\mu }$ and only $\boldsymbol{e}_{0}$ contributes to the 4-velocity in the
Finsler bundle, $\mathbf{u}=\boldsymbol{e}_{0}.$ But, what about the $S_{\mu
i}^{\circ }$ terms? Their contribution is not immediately identifiable with
anything in the paradigm, except possibly for what goes by the name of dark
matter. Hence our choice of the name dark matter for what we are about to
say in this section. Thus let us suppose that we again proceed as in
sections 3 and 4, but with the full $\Omega ^{0}$.

In order not to get lost in manipulating a large number of terms, we shall
keep track of the combinations of terms that become zero by virtue of the
autoparallel conditions $\sigma ^{i}=0$ and $\omega _{0}^{i}=0$. \ We can
express the $\omega _{0}^{i}$ as 
\begin{equation}
\omega _{0}^{i}=a^{i}dt+b_{l}^{i}dx^{l}+c_{m}^{i}du^{m},
\end{equation}%
Before using this in the term $-S_{\text{ \ }\mu i}^{0}\omega ^{\mu }\wedge
\omega _{0}^{i}$ of the torsion, we write it as 
\begin{equation}
...=-S_{\text{ \ }0i}^{0}\omega ^{0}\wedge \omega _{0}^{i}-S_{\text{ \ }%
li}^{0}\omega ^{l}\wedge \omega _{0}^{i}=-S_{\text{ \ }0i}^{0}Ldt\wedge
\omega _{0}^{i}-S_{\text{ \ }0i}^{0}\Lambda _{r}\sigma ^{r}\wedge \omega
_{0}^{i}-S_{\text{ \ }li}^{0}A_{m}^{l}\sigma ^{m}\wedge \omega _{0}^{i}.
\end{equation}%
The last two terms on the right hand side of (29) will contribute to the new 
$\alpha _{i}$ by virtue of the $\sigma $ factors in them. When we then make $%
\omega _{0}^{i}$ equal to zero in the expression for $\alpha _{i}$, they
vanish and thus fail to contribute to the equation of the autoparallels.
Hence, we have to take care of just the first term on the right hand side of
(28), where we use%
\begin{equation}
S_{\text{ \ }0i}^{0}Ldt\wedge (a^{i}dt+b_{l}^{i}dx^{l}+c_{m}^{i}du^{m})=-S_{%
\text{ }0l}^{0}Lb_{i}^{l}dt\wedge \sigma ^{i}-S_{\text{ \ }%
0m}^{0}Lc_{i}^{m}dt\wedge du^{i}),
\end{equation}%
which fits the pattern 
\begin{equation}
\ldots (\qquad )_{l}\wedge \sigma ^{l}+(\ldots )_{i}\text{ }dt\wedge
du^{i}=0.
\end{equation}%
By virtue of the contribution by the last term in (30), we shall have%
\begin{equation}
\Lambda _{i}=L_{.\,i}+S_{0m}^{0}Lc_{m}^{i}
\end{equation}%
instead of $\Lambda _{i}=L_{.\,i}$ The $dL_{.\,i}$ in equation (26) will be
replaced with 
\begin{equation}
dL_{.i}+d(S_{0m}^{0}Lc_{i}^{m}).
\end{equation}%
Hence, the acceleration will emerge not only from $dL_{.i}$, but also from
the dependence of $S_{0m}^{0}Lc_{i}^{m}$ on the $u^{l}.$

The Finsler bundle may not be the ideal way of representing the potential
physical implications that we have uncovered through its use. In a much
later paper, we shall advocate a canonical Kaluza-Klein space without
compactification, where a ``unit vector'' for the fifth dimension embodies
the velocity coordinates. This is not an ad hoc structure. Although he did
not stop at discussing it, Cartan showed that differential geometry is ONLY
a theory of moving frames, not of particles and frames \cite{R8}, pp. 7-8.
That Kaluza-Klein structure, which has in five dimensions a null metric
equivalent to the standard metric, brings in particles at par with frames.

The additional terms that we have just found may be appropriate for
situations like, for example, a current in a metal, which is a quantum
mechanical system. More importantly, the additional acceleration term is of
a nature completely different from the standard one. If this term is in the
right track, there is no hope for velocity curves in galaxies to be
explained by tinkering with the force terms, like is the case with the
postulation of dark matter. It may not be just a ``matter of matter'',
whether dark or not, but also of the format itself of the equation of
motion. We shall nevertheless use the term dark matter sector or simply dark
sector to refer to anything that transcends the form and not just the force
terms in the equation of the motion.

\section{The mathematics of localized energy-momentum: Cartan's differential
invariants}

When one gets used to compute almost exclusively with differential forms, it
is tempting to start a diatribe against the use of energy-momentum tensors.
Tensors blunt the view of Einstein's equations as composed of elements of
the same nature, as opposed to the present view of the Einstein equation
where the Einstein tensor and the energy-momentum tensors belong to
different sides of the equation. All of them are 4-currents, and those
equations then state that a sum of 4-currents equals zero. That is a
perspective different from ``sum of tensors equal zero'', since the term
tensor only speaks of transformation properties, regardless of their
physical significance.

In this section, we speak of scalar-valued differential forms, as
preparation for consideration of vector-valued ones in section 3. In both
cases the argument is very transparent and should be compared with, for
example, section 6 of the first chapter of Landau \& Lifshitz's ``The
Classical Theory of Fields'' \cite{R51}. The last of the 11 pages of that
section look very ad hoc, which is uncharacteristic in those authors, whose
arguments are usually very natural. The section is titled ``Quadrivectors'',
which is strange given that those pages are about integrands. They try to
explain the nature of integrands as antisymmetric tensors, which is a poor
way of explaining that one is dealing with differential $r$-forms as
functions of $r$-surfaces regardless of dimensionality of the manifold. On
the other hand, the tensor product of two antisymmetric tensors is not in
general another antisymmetric tensor. So,\ differential forms look like
antisymmetric tensors but are not, since they are not members of tensor
algebra but of exterior algebra. Exterior algebra is not a subalgebra but a
quotient algebra.

Back to energy-momentum, i.e. that which is conserved under space time
translations. Is it a vector or a differential form? When dealing with
particles, energy-momentum appears to be a tangent 4-vector, since so are
velocity and $d\boldsymbol{P}/d\tau $, where $d\boldsymbol{P}$ is the line
element and $\tau $ is propertime. Other times, it is a differential 3-form
with $dt$ as one of the factors, as in fluid mechanics \cite{R9}. Let $\rho $
be the density of a fluid. We should use the term density to refer to the $%
\rho (x,t)dx^{1}\wedge dx^{2}\wedge dx^{3}$ differential 3-form, since its
integration at constant time is its evaluation as a function of volumes. If
this integral does not depend on time, we have a conserved quantity. The
integral is then called a Poincar\'{e} integral invariant \cite{R9}.

The differential form density can be generalized, meaning that we could have
a ``more complete form'', one whose integration will be independent of
whether it is computed at constant time or not. Cartan showed how to obtain
such a generalization \cite{R9}. One replaces $dx^{i}$ with $dx^{i}-u^{i}dt$
in $\rho dx\wedge dx^{2}\wedge dx^{3}$, which yields 
\begin{equation}
\rho dx^{1}\wedge dx^{2}\wedge dx^{3}-\rho u^{1}dt\wedge dx^{2}\wedge
dx^{3}-\rho u^{2}dt\wedge dx^{3}\wedge dx^{1}-\rho u^{3}dt\wedge
dx^{1}\wedge dx^{2}.  \tag{34}
\end{equation}%
This differential form is said to be a Cartan invariant differential form.
Its components are $\rho (1,-u^{i})$. If we had a density of kinetic energy
rather than one of rest mass or charge, the components of the differential
form would be 
\begin{equation}
\rho \gamma ^{-1}(1,-u^{i})  \tag{35}
\end{equation}%
(where $\gamma =(1-v^{2})^{1/2}$, with $c=1$), i.e. the components of the
kinetic energy-momentum per unit volume. So, they may be viewed as not only
components of a tangent vector, but also of a differential form. The tensor
calculus is very opaque to deal with these issues. It hides the fact that
the first index in energy-momentum tensors, $T^{\mu \nu }$, is a valuedness
index and the second one is a differential form index, specifically the one
missing in monomial differential 3-forms (in dimension four!). More
generally, consider any scalar-valued differential 3-form in spacetime 
\begin{equation}
j=j^{0}dx^{1}\wedge dx^{2}\wedge dx^{3}-j^{i}dt\wedge dx^{j}\wedge dx^{k}, 
\tag{36}
\end{equation}%
with sum over cyclic permutations of $(i,j,k)=1,2,3$. Assume that $dj=0$. We
then have 
\begin{equation}
0=\left( \frac{\partial j^{0}}{\partial t}+\frac{\partial j^{1}}{\partial
x^{1}}+\frac{\partial j^{2}}{\partial x^{2}}+\frac{\partial j^{3}}{\partial
x^{3}}\right) dt\wedge dx^{1}\wedge dx^{2}\wedge dx^{3}.  \tag{37}
\end{equation}%
The usual but not sufficiently transparent way of writing (4) is 
\begin{equation}
\frac{\partial \rho }{\partial t}+\mbox{div}\,\boldsymbol{j}=0.  \tag{38}
\end{equation}%
The appropriate way of course is 
\begin{equation}
0=dj=d(j^{0}dx^{123}-j^{i}dx^{0jk}),  \tag{39}
\end{equation}%
which makes $-j^{i}dx^{0jk}$ the components of a flux of some scalar-valued
quantity whose density is given by $j^{0}dx^{123}.$ E. K\"{a}hler used this
observation to obtain positrons with the same sign of energy as electrons in
the quantum mechanics that is a concomitant of his calculus \cite{R48}. In
the same spirit, and helped by the geometrization of electrodynamics, the
view of the conservation law in terms of differential forms will show us
where a true, local concept of gravitational energy-momentum ``tensor'' has
been hiding.

Following K\"{a}hler, the conservation of a quantity that comes in two
forms, like charge does, is given by 
\begin{equation}
\frac{\partial \rho ^{(1)}}{\partial t}+\sum_{i}j_{i}^{(1)}dt\wedge
dx^{j}\wedge dx^{k}+\frac{\partial \rho ^{(2)}}{\partial t}%
+\sum_{i}j_{i}^{(2)}dt\wedge dx^{j}\wedge dx^{k}=0.  \tag{40}
\end{equation}%
This idea can be extended to conserved quantities that come in more than two
varieties, like energy-momentum does.

\section{The mathematics of energy-momentum ``tensors''}

Assume that we are confronted with the computation of some directional
quantity distributed over ---to make things simple--- a spherical domain.
If, in addition, the distribution is spherically symmetric, a choice of
coordinates can simplify the form of the integrand. But the vector-valued
integrand (vector-valued in the same that we are dealing with components of
a vector quantity) will have to be projected over constant directions for
its integration, since the directions of vectors tangent to coordinate lines
change from point to point. In arbitrary manifolds, constant directions\ do
not even exist in general and, therefore, vector-valued integrations do not
make sense, ab initio. In particular, they do not make sense in modern
relativistic gravitation theory, where the Euclidean connection of spacetime
is assumed to be Levi-Civita's, which is not teleparallel.

It is important to realize that the original theory of general relativity
did not involve any connection at all. The concept of connection only
emerged in 1917 with the discovery of what nowadays is known as the LC
connection \cite{R52}.

One may retort that such integrations appear to be working very well in
general relativity. Indeed they do. But is the assumption of LC\ connection
being used in the process, or is it simply that we only see the ``metric
part'' of the connection because we are unable to identify the other,
torsional part? We shall see in section 6 a most instructive example of this
mechanism. So, it might well be that general relativity works because of the
underlying teleparallel connection.

Because of the existence of constant frame fields under the assumption of
TP, it is legitimate to then speak of vector-valued quantities and,
therefore, of energy-momentum tensors, except, once again, that we should
not be speaking about such tensors but about vector-valued differential
3-forms.

The relation between these two concepts will now be explained in detail. Let 
$\omega ^{\mu \nu \lambda }$ denote $\omega ^{\mu }\wedge \omega ^{\nu
}\wedge \omega ^{\lambda }$, to be also written as $\omega ^{\mu \nu \lambda
}$. Consider the vector-valued differential 3-form $\boldsymbol{G}$ given by 
\begin{equation}
\boldsymbol{G}\equiv G_{\mu \nu \pi }^{\alpha }(\omega ^{\mu \nu \lambda })%
\boldsymbol{e}_{\alpha },  \tag{41}
\end{equation}%
where the parenthesis around $\omega ^{\mu \nu \lambda }$ indicates that we
are summing over a basis of differential 3-forms, not over repeated indices.
If $d\boldsymbol{G}$ vanishes on a region and we assume $TP$, a vector
valued quantity is conserved. In constant frame fields, the equation $d%
\boldsymbol{G}=0$ becomes 
\begin{equation}
d[G_{\mu \nu \pi }^{\alpha }(\omega ^{\mu \nu \lambda })]=0,  \tag{42}
\end{equation}%
i.e. four conservation laws, one for each of the four values of the
superscript in those components.

Let $z$ denote the unit differential 4-form in the K\"{a}hler algebra, i.e.
the Clifford algebra of scalar-valued differential forms. In other words, it
is the associative algebra defined by 
\begin{equation}
dx^{\mu }\vee dx^{\nu }+dx^{\nu }\vee dx^{\mu }=2g^{\mu \nu }.  \tag{43}
\end{equation}%
In the canonical bases defined by a Lorentzian metric, we instead have 
\begin{equation}
\omega ^{\mu }\vee \omega ^{\nu }+\omega ^{\nu }\vee \omega ^{\mu }=2\eta
^{\mu \nu },  \tag{44}
\end{equation}%
where $\eta ^{\mu \nu }$ is diagonal $(1,-1,-1,-1)$ or $(-1,1,1,1).$ Let
juxtaposition of differential forms denote Clifford product. Clifford
multiplication of a differential form by $z$ is equivalent, in the tensor
calculus, to contraction with the Levi-Civita tensor. Hence $\boldsymbol{G}z$
is a vector-valued differential 1-form 
\begin{equation}
\boldsymbol{G}z=G_{\nu }^{\text{ }\mu }\omega ^{\nu }\mathbf{e}_{\mu
}=G^{\nu \mu }\omega _{\mu }\mathbf{e}_{\nu }.  \tag{45}
\end{equation}%
The $G_{\nu }^{\mu }$ behave like the components of a second rank tensor. On
account of Eqs. (41) and (45), we have 
\begin{equation}
G^{\alpha \mu }\omega _{\mu }=G_{\lambda \nu \pi }^{\alpha }(\omega
^{\lambda \nu \pi })z.  \tag{46}
\end{equation}%
Therefore, $G^{\alpha \mu }=G_{\lambda \nu \pi }^{\alpha }$ since $\omega
_{\mu }=\omega ^{\lambda \nu \pi }z$, all four indices being different. Care
has to be taken with the conventions adopted, like in defining $z$, as they
may bring about signs $-1$.

We define differential 3-forms $G^{\alpha }$ by means of $\boldsymbol{G}%
=G^{\alpha }\boldsymbol{e}_{\alpha }.$ Then, 
\begin{equation}
G^{\alpha }\equiv G_{\mu \nu \pi }^{\alpha }(\omega ^{\mu \nu \lambda }). 
\tag{47}
\end{equation}%
Knowing $G^{\alpha }$, whose form is easy to remember in Einstein's case
(see below), we would have to multiply by $z$, as per Eqs. (45)-(46) in
order to get the Einstein tensor.

In terms of the curvature, the components as a vector of the Einstein
vector-valued differential 3-form are given by \cite{R8}, \cite{JGV52} 
\begin{align}
& G^{0}=\omega ^{1}\wedge \underline{\Omega }^{23}+\omega ^{2}\wedge 
\underline{\Omega }^{31}+\omega ^{3}\wedge \underline{\Omega }^{12}, 
\tag{48} \\
& G^{i}=\omega ^{0}\wedge \underline{\Omega }^{jk}+\omega ^{j}\wedge 
\underline{\Omega }^{k0}+\omega ^{k}\wedge \underline{\Omega }^{0j}. 
\tag{49}
\end{align}%
In (49), we mean the three cyclic permutations of (1,2,3). Notice that the
index on the left is the missing index on the right.

In the interest of brevity of notation, we shall sometimes use the term
Einstein tensor when there is no doubt that we are referring to the
differential forms (48)-(49).

\section{Of general relativity and the connection of spacetime}

In a reputed book on general relativity, we read that ``anybody who looks
for a magical formula for local gravitational energy-momentum is looking for
the right answer to the wrong question''. And, in the same paragraph, the
closing statement reads ``One can always find in any given locality a frame
of reference in which all local gravitational fields (all Christoffel
symbols; all $\Gamma _{\mu \nu }^{\alpha }$) disappear. No $\Gamma $ means
no gravitational field and no local gravitational field means no local
gravitational energy-momentum.''

One can always find a frame of reference in which all the $\Gamma _{\mu \nu
}^{\alpha }$ vanish at a given point, but not in its neighborhood, or else
the curvature at the point would be zero. So, the $\Gamma $'s are
irrelevant; if the determining factor were their annulment or not, there is
local gravitational energy-momentum in the neighborhood of the chosen point.
So the argument is self-defeating.

General relativists believe that the (pseudo) Euclidean connection of
spacetime is the Levi-Civita connection. But this so by historical accident
since this connection was born in 1917. In that year, Levi-Civita (but also
Schouten, according to Cartan, and Hessenberg, according to Abraham Pais)
had realized that one could use the Christoffel symbols to perform parallel
transport of vectors. Such transport became a path-dependent imitation of
equality of tangent vectors at different point of a differentiable manifold.
The affine concept of connection was born with the new, additional role
assigned to those symbols. It was an affine role, not a metric one, since
parallel transport can also be defined in the absence of a metric, as E.
Cartan showed in 1922 \cite{R11} \ See also his series of letters in 1922 to
the French academy, specially \cite{R10}, where he wrote and explained that
``the metric does not contain all the geometric reality of a space''.

The Levi-Civita (LC) connection was adopted by physics within a few years.
One can find it in Pauli's famous 1921 book on GR \cite{R57}. At that point,
there was no other candidate connection for adoption other than alternatives
by Pauli and Eddington, readily discarded. But there was not yet a general
theory of connections to provide guidance. Since there was no LC connection
at the birth in 1915 of $GR$, there were two epochs in this theory, say
before and after the adoption of the LC\ connection. This fact is seriously
overlooked by the physics paradigm.

We insist on the fact that the presence of Christoffel symbols, $\underline{%
\Gamma }_{\mu \text{ \ }\lambda }^{\text{ \ }\nu }$, does not imply that a
theory contains a connection, even through those symbols are the components
in coordinate bases of the a differential 1-form $\alpha _{\mu }^{\nu }$%
\begin{equation}
\alpha _{\mu }^{\nu }=\underline{\Gamma }_{\mu \text{ \ }\lambda }^{\text{ \ 
}\nu }dx^{\lambda },  \tag{50}
\end{equation}%
of relevance on differentiable manifolds endowed with a metric. Until 1917,
these symbols had nothing to do with comparing vectors at a distance, or for
transporting tangent vectors for that matter. So, up to that point, there
was no reason to refer to $\alpha _{\mu }^{\nu }$ as a connection. Nothing
was being connected. The $\underline{\Gamma }_{\mu \text{ \ }\lambda }^{%
\text{ \ }\nu }$ had constituted simply useful intermediate quantities in
the process of going from the metric to the Riemannian curvature, which was
viewed just as a set of quantities before the tensor calculus was
formulated. The annulment of those quantities was a necessary and sufficient
condition for two symmetric, quadratic differential forms to be transformed
one into the other by a coordinate transformation. And they were useful in
the compact formulation of the equations of curves of stationary length. But
when we have a Euclidean or pseudo-Euclidean connection, i.e. what is
usually called a metric compatible affine connection (affine it is not!),
the parallel transport need not be given by $\alpha _{\mu }^{\nu }$. It is,
therefore, not legitimate to refer to $\alpha _{\mu }^{\nu }$ as the LC
connection, except in cases when we choose it to play the ``affine role'' of
defining the parallel transport, i.e. choosing for $\omega _{\mu }^{\nu }$
the $\alpha _{\mu }^{\nu }$ itself.

In the late 1920's and without referring to the aforementioned two epochs,
Einstein argued that something was missing in general relativity \cite{R39}.
His ``instinct'' told him that the solution had to be in TP, which he called
absolute parallelism. Years later he would give it up, owing to his total
lack of success at finding any physics in it. Cartan knew how one had to
exploit TP. Retrospectively speaking, his unheeded recommendations to
Einstein were totally right. But Cartan would not do the required work on
Finsler geometry until the mid 1930's, neither did he returned to the
subject of physical unification, nor did Einstein return to TP.
Retrospectively, it would have been better for physics if no affine role had
been assigned to $\alpha _{\mu }^{\nu }$. But who knew then that there was
an infinite number of connections containing the same metric contents as the
old Riemannian geometry, their difference lying in their affine contents?
The remainder of this paper will elaborate on this point.

To conclude our consideration on the evolution of Riemannian geometry and
its impact on the theory of relativistic gravitation, let us say that
manifolds endowed with Euclidean connections do have two curvatures, the
Riemannian one, 
\begin{equation}
\underline{\Omega }_{\mu }^{\nu }=d\alpha _{\mu }^{\nu }-\alpha _{\mu
}^{\lambda }\wedge \alpha _{\lambda }^{\nu },  \tag{51}
\end{equation}%
and the Euclidean (or pseudo-Euclidean) curvature 
\begin{equation}
\Omega _{\mu }^{\nu }=d\omega _{\mu }^{\nu }-\omega _{\mu }^{\lambda }\wedge
\omega _{\lambda }^{\nu }.  \tag{52}
\end{equation}

\section{The first pair of Maxwell's equations}

The first Bianchi identity reads 
\begin{equation}
d\Omega ^{\mu }=-\Omega ^{\nu }\wedge \omega _{\nu }^{\mu }+\omega ^{\nu
}\wedge \Omega _{\nu }^{\mu }.  \tag{53}
\end{equation}%
In $TP$, the last term vanishes. The identification of Maxwell's equations
within geometry is facilitated by avoiding the complications of a non-flat
metric. Hence, if we assume that the metric is sufficiently weak, we choose
coordinate bases of differential forms in terms of which the LC\
differential form $\alpha _{\mu }^{\text{ }\nu }$ vanishes, i.e. Cartesian.
We then have, with $\beta _{\mu }^{\nu }$ defined as $\omega _{\mu }^{\nu
}-\alpha _{\mu }^{\nu }$ 
\begin{equation}
d\Omega ^{\mu }+\Omega ^{\nu }\wedge \beta _{\nu }^{\mu }=0.  \tag{54}
\end{equation}%
The components of $\beta _{\nu }^{\mu }$ are linear combinations of the
components of the torsion. Hence the last term on the left of (54) is
quadratic in the torsion. We make ours the words that Einstein pronounced in
his letter to Cartan of December 27-28 of 1929 \cite{R29}: ``But no
reasonable person believes that Maxwell's equations can hold rigorously.
They are, in suitable cases, first approximations for weak fields''. The
linear approximation of (54) is $d\Omega ^{\mu }=0.$ The equation $d\Omega
^{0}=0$ is the first pair of Maxwell's equations, $dF=0$, after the
identification we made of $\Omega ^{0}$ with $F$ for electromagnetic
torsions.

Once we have got this result on the flat metric, let us observe that the
equation (53) with $\Omega _{\nu }^{\mu }=0$ can be written as 
\begin{equation}
d(\Omega ^{\mu }\boldsymbol{e}_{\mu })=d\Omega ^{\mu }+\Omega ^{\nu }\wedge
\omega _{\nu }^{\mu }=0,  \tag{55}
\end{equation}%
which coincides with (54) for flat metric. Using as before $\omega _{\mu
}^{\nu }=\alpha _{\mu }^{\nu }+\beta _{\mu }^{\nu }$, using semicolon for
covariant differentiation with respect to the metric and expressing $\beta
_{\mu }^{\nu }$ in terms of the torsion, we get the equation that Cartan
proposed to Einstein in letter of December 3, 1929: 
\begin{equation}
\Lambda _{\alpha \beta ;\gamma }^{\mu }+\Lambda _{\beta \gamma ;\alpha
}^{\mu }+\Lambda _{\gamma ^{\alpha ;\beta }}^{\mu }+\Lambda _{\alpha \beta
}^{p}\Lambda _{\rho \gamma }^{\mu }+\Lambda _{\beta \gamma }^{\rho }\Lambda
_{p\alpha }^{\mu }+\Lambda _{r\alpha }^{p}\Lambda _{p\beta }^{\mu }=0. 
\tag{56}
\end{equation}
Cartan did not use the term first Bianchi identity in proposing (56), but he
stated that \textit{for the physicist}, they are not identities but express
physical laws of nature \cite{R29} (emphasis in the original).

The equation $d\Omega ^{0}=0$ is invariant in Finsler geometry ---not in
Riemannian geometry with torsion--- because the group in the fibers is $O(3)$%
, not the Lorentz group. In liftings to the Finsler bundle of connections on
the standard bundle of manifolds endowed with the Lorentz metric, boosts
emerge in the integration of $\omega _{0}^{i}\boldsymbol{e}_{i}$ between two
points on Finslerian sections. Recall that $TP$ admits constant frame fields
and thus allows for path independent integration of $\omega _{0}^{i}%
\boldsymbol{e}_{i}$.

\section{The second equation of structure in Finslerian teleparallelism and
the full geometrization of Einstein's equations}

Teleparallelism (TP) means that it is possible to establish a relation of
equality of vectors (i.e. path independent) among tangent vectors at any two
points of a connected manifold. In terms of differential equations, it
requires that the Lorentzian curvature 
\begin{equation}
\Omega _{\mu }^{\nu }\equiv d\omega _{\mu }^{\nu }-\omega _{\mu }^{\lambda
}\wedge \omega _{\lambda }^{\nu }  \tag{57}
\end{equation}%
be zero. The Riemannian curvature, 
\begin{equation}
\underline{\Omega }_{\mu }^{\nu }\equiv d\alpha _{\mu }^{\nu }-\alpha _{\mu
}^{\lambda }\wedge \alpha _{\lambda }^{\nu }  \tag{58}
\end{equation}%
remains available for use in TP, the $\alpha _{\mu }^{\text{ }\nu }$ being
the solution of the system 
\begin{equation}
d\omega ^{\mu }=\omega ^{\nu }\wedge \alpha _{\nu }^{\text{ }\mu },\quad \ \
\ \ \ \ \ \ \ \ \ \ \ \ \alpha _{\mu \nu }+\alpha _{\nu \mu }=dg_{\mu \nu },
\tag{59}
\end{equation}%
with $g_{\mu \nu }\equiv \mathbf{e}_{\mu }\cdot \mathbf{e}_{\nu }$.

Cartan and Einstein corresponded with each other on TP from May of 1929 to
May 1932 \cite{R29}, by which time Einstein had already abandoned his
attempt at whatever he had expected to achieve with it (unification of
interactions?, geometrization of electrodynamics?, a more classical looking
replacement for quantum theory?, all of that?). Cartan suggested to him
equations that should go into any system that exploits TP for physical
theory building. A most important point of discussion that recurred time and
again in their correspondence was whether there should be a role for the
Ricci tensor in that system. Einstein's answer was in the negative. Cartan
on the other hand argued that for the physics to be deterministic one would
have to bring in the Riemannian curvature, like through Ricci or Einstein
tensors.

With caveats, which have to do mainly with the fact that quantum mechanics
has the last word, we shall adopt here Einstein's thesis of logical
homogeneity of differential geometry and theoretical physics. A strict
interpretation of this thesis means that the field equations of the theory
should be the equations of structure and Bianchi identities.

In TP, the second equation of structure yields 
\begin{equation}
0=d\omega _{\mu }^{\nu }-\omega _{\mu }^{\lambda }\wedge \omega _{\lambda
}^{\nu }=d(\alpha _{\mu }^{\nu }+\beta _{\mu }^{\nu })-(\alpha _{\mu
}^{\lambda }+\beta _{\mu }^{\lambda })\wedge (\alpha _{\lambda }^{\nu
}+\beta _{\lambda }^{\nu }).  \tag{60}
\end{equation}%
We reorganize this to obtain 
\begin{equation}
\underline{\Omega }_{\mu }^{\nu }\equiv d\alpha _{\mu }^{\nu }-\alpha _{\mu
}^{\lambda }\wedge \alpha _{\lambda }^{\nu }=\beta _{\mu }^{\lambda }\wedge
\beta _{\lambda }^{\nu }-(d\beta _{\mu }^{\nu }+\alpha _{\mu }^{\lambda
}\wedge \beta _{\lambda }^{\nu }+\beta _{\mu }^{\lambda }\wedge \alpha
_{\lambda }^{\nu }).  \tag{61}
\end{equation}%
If we replace $\alpha _{\pi }^{\text{ }\rho }$ with $\omega _{\pi }^{\text{ }%
\rho }-\beta _{\pi }^{\text{ }\rho }$, (61) becomes%
\begin{equation}
\underline{\Omega }_{\mu }^{\nu }\equiv d\alpha _{\mu }^{\nu }-\alpha _{\mu
}^{\lambda }\wedge \alpha _{\lambda }^{\nu }=-\beta _{\mu }^{\lambda }\wedge
\beta _{\lambda }^{\nu }-(d\beta _{\mu }^{\nu }+\omega _{\mu }^{\lambda
}\wedge \beta _{\lambda }^{\nu }-\omega _{\lambda }^{\nu }\wedge \beta _{\mu
}^{\lambda }).  \tag{62}
\end{equation}
We shall later see why (62) is preferable over (61).

Define $\underline{\mho }$ and $\boldsymbol{\beta}$ as%
\begin{equation}
\underline{\mho }=\underline{\Omega }_{\mu }^{\text{ }\nu }\boldsymbol{e}%
^{\mu }\wedge \boldsymbol{e}_{\nu },  \tag{63}
\end{equation}%
and 
\begin{equation}
\mathbf{\beta }\equiv \beta _{\mu }^{\text{ }\nu }\boldsymbol{e}^{\mu
}\wedge \boldsymbol{e}_{\nu }.  \tag{64}
\end{equation}%
Then, from (61)-(64), one gets \ 
\begin{equation}
\underline{\mho }=-(\beta _{\mu }^{\lambda }\wedge \beta _{\lambda }^{\nu })%
\boldsymbol{e}^{\mu }\wedge \boldsymbol{e}_{\nu }-d\boldsymbol{\beta}. 
\tag{65}
\end{equation}%
The Einstein contraction of $\underline{\mho }$ is the Einstein tensor.
Hence the contraction of (65) is a fully geometrized Einstein equation.
Through the relation of the contorsion to the torsion and in turn to the
electromagnetic field, and the choice of torsions with $\Omega ^{i}=0,$ $S_{%
\text{ \ }\nu l}^{\mu }=0$, the contraction of (65) starts to look physical.
In the next section, we shall test whether the right electromagnetic
energy-momentum is represented in the contraction of the quadratic term in
the torsion. Assume for the moment that it is. Then $\Omega ^{i}\neq 0$ and $%
S_{\text{ \ }\nu l}^{\mu }\neq 0$ allows in principle for the classical
representation of much more physics. But, in the last instance, the world is
quantum mechanical. Hence the role of $\Omega ^{i}$ and $S_{\text{ \ }\nu
l}^{\mu }$ in the classical representation might only emerge from the
quantum mechanical sector of a hypothetical theory which comprises and
extends present differential geometry (this extension will be explained in a
paper under development).

The most important implication of Einstein's contraction of (65) has to do
with the fact that it finally solves the problem of where the gravitational
energy-momentum current lies. Start by rewriting (65) as 
\begin{equation}
\underline{\mho }+(\beta _{\mu }^{\lambda }\wedge \beta _{\lambda }^{\nu })%
\boldsymbol{e}^{\mu }\wedge \boldsymbol{e}_{\nu }+d\boldsymbol{\beta}=0. 
\tag{66}
\end{equation}%
This is a sum of bivector-valued differential 2-forms. Define 
\begin{equation}
\boldsymbol{Z}\equiv \boldsymbol{e}^{0}\boldsymbol{e}^{1}\boldsymbol{e}^{2}%
\boldsymbol{e}^{3},  \tag{67}
\end{equation}%
where juxtaposition means Clifford product and where the basis of tangent
vectors is pseudo-orthonormal. Let $(\wedge ,\cdot )$ represent exterior
product in the algebra of differential forms, and dot product in the tangent
Clifford algebra. The Einstein equations then are%
\begin{equation}
0=d\mathbf{P}(\wedge ,\cdot )\mho \mathbf{Z}=\omega ^{\mu }\mathbf{e}_{\mu }%
\text{ }(\wedge ,\cdot )\text{ }[\underline{\mho }+(\beta _{\mu }^{\lambda
}\wedge \beta _{\lambda }^{\nu })\boldsymbol{e}^{\mu }\wedge \boldsymbol{e}%
_{\nu }+d\boldsymbol{\beta}]\boldsymbol{Z}.  \tag{68}
\end{equation}%
The right hand side can be seen as the sum of three vector-valued
differential 3-forms, the first one of them being the Einstein current 
\begin{equation}
\omega ^{\mu }\boldsymbol{e}_{\mu }(\wedge ,\cdot )\underline{\mho }\mathbf{Z%
},  \tag{69}
\end{equation}%
which is equivalent to specifying the Einstein tensor. Its components were
given as equations (48) and (49).

The way in which the Einstein ``tensor'' has emerged ---in the same way as
the other two in the second equation of structure--- strongly suggests that
we have to view (68) as the statement that a sum of energy-momentum currents
is zero, and that, therefore, energy-momentum is conserved. Needless to say
that (69), i.e. the first term on the right hand side of (68) can only be
gravitational, since it only depends on the metric. The second term should
be seen as ``generalized'' electromagnetic energy-momentum tensor (meaning
here electromagnetic and beyond). The third term appears to combine the
gravitational interaction with those buried in the torsion. Let us simply
say at this point that, by elimination, the standard cosmological term might
be a special case of $d\boldsymbol{\beta}$. Before we go into this, let us
make a remark on the relation the first Bianchi identity to the second
equation of structure, both in TP.

The explicit form of the difference between the first Bianchi identity and
(53) is given by the equations 
\begin{equation}
\omega ^{1}\wedge \Omega _{1}^{0}+\omega ^{2}\wedge \Omega _{2}^{0}+\omega
^{3}\wedge \Omega _{3}^{0}=0,  \tag{70}
\end{equation}%
\begin{equation}
\omega ^{0}\wedge \Omega _{0}^{i}+\omega ^{j}\wedge \Omega _{j}^{\text{ }%
i}+\omega ^{k}\wedge \Omega _{k}^{\text{ }i}=0,  \tag{71}
\end{equation}%
where (71) is meant for $(i=1,$ $j=2,$ $k=3)$ and its cyclic permutations.
Since the $\omega ^{\nu }$ constitute a basis of differential 1-forms, the
components of the 3-forms in (70)-(71) are the same as the components $%
R_{\mu \text{ }\lambda \rho }^{\text{ }\nu }$ of the curvature 2-forms
themselves, as is the case with the components of the terms in Eqs.
(48)-(49). Hence both sets of equations are linear combinations of the
components $R_{\mu \text{ }\lambda \rho }^{\text{ }\nu }$ of the curvature,
and both contribute in a similar manner towards determining the $R_{\mu 
\text{ }\lambda \rho }^{\text{ }\nu }$, though not actually determining
them. More equations would be needed for that.

Assume that we choose for $\Omega _{\mu }^{\nu }$ in (70)-(71) \ the
Riemannian curvature $\underline{\Omega }_{\mu }^{\text{ }\nu }$. We compare
these expressions with the components as a vector of the Einstein current,
which were given in (14)-(15) but after replacing the Riemannian curvature
with general Euclidean or pseudo-Euclidean curvature. One then sees that, in
the so modified system (14)-(15), one has the exterior product contractions
of four sets of three $\omega ^{\mu }$'s with different combinations of the
components of the curvature. This is like in (51)-(52), but for a different
choice of component of the same curvature. These are the only two covariant
options to obtain differential 3-forms from the curvature 2-form.

The foregoing equations amount to 
\begin{equation}
(\omega ^{\nu }\wedge \Omega _{\nu }^{\text{ }\mu })\boldsymbol{e}_{\mu }=0,
\tag{72}
\end{equation}%
whereas the Einstein's equation is 
\begin{equation}
\omega ^{\mu }\boldsymbol{e}_{\mu }(\wedge ,\cdot )\mathbf{\mho }%
\boldsymbol{Z}=0.  \tag{73}
\end{equation}

Physics has grown inductively with the help of phenomenology. The main goal
of physics in general is not the study of physical systems in all their full
complexity, but to first extract in a complex situation the minimum number
of elements that allow one to qualitatively arrive to a reality-conforming
solution of the issues raised for physical systems. That is why the main
physical laws are the way they are. The concept of field is only a few
centuries old. Basic and still relevant concepts like mass, weight, length,
distance,velocity, brightness, acceleration, even lightning, have to do or
at least first had to do with matter. That is why, even in teleparallelism,
we are interested in that the sum of the currents is zero, and not that the
corresponding terms in the expansion of the curvature is zero. The latter is
too remotely connected with the situations that one faces in classical
physics. But the curvature equations would be more significant for
formulating a theory of everything or at least taking us in that direction.

\section{On the cosmological term}

We shall now deal with the ``relation'' of $d\boldsymbol{\beta}$ to the
cosmological term. TP implies the existence of frames fields where $\omega
_{\nu }^{\text{ }\mu }$, which does not transform like the components of a
tensor, vanishes. Thus $\beta _{\mu }^{\lambda }=-\alpha _{\mu }^{\lambda }$
in those frame fields, namely the fields where matter at the highest levels
of structure is at ``relative rest''. Statements like the ones just made
about the relation between $\beta _{\mu }^{\lambda }$ and $\alpha _{\mu
}^{\lambda }$ have to be seen in the light of the fact that torsion fields
(thus also $\beta _{\mu }^{\lambda }$) are not directly the fields of the
interactions, but only, at this stage, through factors like $q/m$ and $G$
and possibly others.

TP justifies performing integrations of vector- and bivector--valued
quantities, like vector-valued currents (read tensors if you will) and
curvatures. Adding (read integrating) their components when the basis is not
the same at different points ---only TP allows their equality--- means
nothing at all, even in flat spaces. It is in those fields that $\beta _{\mu
}^{\lambda }$ equals $-\alpha _{\mu }^{\lambda }$, regardless of whether
these differential forms are large or small. Two different $\beta $'s can
give rise to the same gravitational field, say outside two balls of matter.
But the torsion, and thus $\beta _{\mu }^{\lambda }$, has more independent
components than the metric, which determines $\alpha _{\mu }^{\lambda }$.

In trying to obtain the structure of the universe in terms of the
''observed'' matter, whether ordinary or dark, we take as input its
distribution, seen or inferred. For that purpose, one has to use an equation
such as (61) or (62) ---actually (62) better than (61) for it is $\omega
_{\nu }^{\text{ }\mu }$, not $\alpha _{\mu }^{\lambda }$ that can be made
equal to zero. It is then the quadratic term in $\beta $ that has to be
associated with anything that we call at present matter or radiation, and it
is $d\boldsymbol{\beta}$ that needs to be interpreted. The equation $\omega
_{\nu }^{\text{ }\mu }=0$, and thus $\beta _{\mu }^{\lambda }=-\alpha _{\mu
}^{\lambda }$, applies only at the level of formulating the basic equations,
and not even there if we accept any sort of fluctuations as invoked in very
early times in the light of the cosmological model. Much less does it apply
when structure has developed in the evolution of the universe.

In view of the above, it is then not to be expected that $\beta _{\nu }^{%
\text{ }\mu }$ and $\alpha _{\mu }^{\lambda }$ will be each other's opposite
except as averages over regions much larger than superclusters of galaxies.
We thus have to go to the Einstein equation, where we neglect the quadratic
term in $\beta _{\mu }^{\lambda }$ in (62).

\begin{equation}
d\alpha _{\mu }^{\nu }-\alpha _{\mu }^{\lambda }\wedge \alpha _{\lambda
}^{\nu }=-(d\beta _{\mu }^{\nu }+\omega _{\mu }^{\lambda }\wedge \beta
_{\lambda }^{\nu }+\beta _{\mu }^{\lambda }\wedge \omega _{\lambda }^{\nu }).
\tag{74}
\end{equation}%
We need not neglect $-\alpha _{\mu }^{\lambda }\wedge \alpha _{\lambda
}^{\nu }$ since, as just argued, we may not do so when actually involving
structure in post-primordial times. It is, however, reasonable to do so for
this stage of the evolution of the universe, since most of the space of the
universe is empty of matter. In that emptiness $\beta _{\mu }^{\lambda
}\backsim -\alpha _{\mu }^{\lambda }$ and $\omega _{\mu }^{\lambda }$ $%
\backsim 0.$ Thus, approximately, 
\begin{equation}
d\alpha _{\mu }^{\nu }=-d\beta _{\mu }^{\nu }.  \tag{75}
\end{equation}%
Of course, this is implied by $\beta _{\mu }^{\lambda }=-\alpha _{\mu
}^{\lambda }$, but not the other way around. (75) implies 
\begin{equation}
\beta _{\mu }^{\nu }=-\alpha _{\mu }^{\nu }+d\gamma _{\mu }^{\nu },  \tag{76}
\end{equation}%
where the $\gamma _{\mu }^{\nu }$ are scalar functions ($0$-forms). Hence,
if the present argument is correct, the cosmological constant is, in TP,
just the tip of an iceberg, but not quite yet.

Indeed, the $\gamma _{\mu }^{\nu }$ will go into the cosmological term in
the Einstein equations where $\beta _{\mu }^{\lambda }\wedge \beta _{\lambda
}^{\nu }$ is replaced in terms of the energy-momentum of matter. But we
would then have $dd\gamma _{\mu }^{\nu }$, which is zero. So, it would seem
that there is no cosmological term after all. But this is a limitation of
general relativity because of a limitation of differential geometry. This
geometry is about the exterior calculus of vector-valued differential forms.
One needs a differential geometry based not on the exterior calculus but on
the K\"{a}hler calculus, i.e. organic sum of the interior and exterior
derivatives. This is clear if we have in mind Helmholtz theorem, where even
for a vector field we need not only the curl but also the divergence. Hence,
the final word for the cosmological problem in TP, as for the
energy-momentum of electrodynamics and for the in depth geometrization of
the second pair of Maxwell's equations will have to wait for the extension
of differential geometry in the direction just indicated. We can advance
this much, however. One day, observations may need more sophistication than
presently required by cosmological models.

\section{On the energy-momentum of electrodynamics}

We now proceed to consider the geometric version of electromagnetic
energy-momentum in the limited context of present differential geometry. We
achieve economy of computations by observing that in the Einstein
vector-valued differential 3-form, (48)-(49), $G^{0}$ is the current for
gravitational energy, and the $G^{i}$ are the currents for the components of
gravitational momentum. For electromagnetic energy, we take the right hand
side of (48) and replace in it $\underline{\Omega }^{jk}$ with $\beta ^{jk}.$
We get:%
\begin{equation}
\omega ^{1}\wedge \beta ^{2\lambda }\wedge \beta _{\lambda }^{3}+\omega
^{2}\wedge \beta ^{3\lambda }\wedge \beta _{\lambda }^{1}+\omega ^{3}\wedge
\beta ^{1\lambda }\wedge \beta _{\lambda }^{2}.  \tag{77}
\end{equation}%
The overall sign can be ignored since it is only after we have obtained the
electromagnetic energy-momentum that the sign is determined, not only for
this energy contribution but for all terms after they have been put on the
same sign of the equation

We solve for $\beta _{\mu }^{\nu }$ the system 
\begin{equation}
\Omega ^{\mu }=-\omega ^{\nu }\wedge \beta _{\nu }^{\mu },\;\ \ \ \ \ \ \ \
\ \ \ \ \ \ \ \ \ \ \ \ \ \ \ \ \ \ \ \ \ \ \ \ \ \ \ \ \ \ \;\beta _{\mu
\nu }+\beta _{\nu \mu }=0,  \tag{78}
\end{equation}%
and obtain 
\begin{equation}
\beta _{\mu \nu }=\beta _{\mu \nu \lambda }\omega ^{\lambda }\text{ \ \ \ \
\ \ \ \ \ \ \ \ \ \ \ }\beta _{\mu \nu \lambda }=\frac{1}{2}(R_{\mu \nu
\lambda }+R_{\nu \lambda \mu }+R_{\lambda \nu \mu }),  \tag{79}
\end{equation}%
which is to be taken to (77):%
\begin{align}
\beta ^{j\lambda }\wedge \beta _{\lambda }^{\text{ }k}& =\beta ^{j\text{ }%
0}\wedge \beta _{0}^{\text{ }k}-\beta ^{ji}\wedge \beta _{i}^{\text{ }%
k}=\beta _{j0}\wedge \beta _{0k}+\beta _{ji}\wedge \beta _{ik}  \notag \\
& =-\beta _{0j}\wedge \beta _{0k}+\beta _{ij}\wedge \beta _{ki},  \tag{80}
\end{align}%
with signature -2. The restriction in (79) to $R_{i\mu \nu }=0,$ yields 
\begin{equation}
\beta _{0i0}=R_{0i0}=-E_{i},\text{ \ \ \ \ \ \ }\beta _{0ij}=\frac{1}{2}%
R_{0ij}=-\frac{1}{2}B_{k},,\text{ \ \ \ \ \ \ }\beta _{i00}=R_{00i}=E_{i,} 
\tag{81}
\end{equation}%
\begin{equation}
\beta _{i0j}=\frac{1}{2}R_{0ji}=\frac{1}{2}B_{k},\text{ \ \ \ \ \ \ \ }\beta
_{ij0}=\frac{1}{2}R_{0ji}=\frac{1}{2}B_{k},\ \ \ \ \ \ \ \beta _{0ii}=\beta
_{ijk}=0,  \tag{82}
\end{equation}%
and%
\begin{equation}
\beta _{0i}=\beta _{0i0}\omega ^{0}+\beta _{0ii}\omega ^{i}+\beta
_{0ij}\omega ^{j}+\beta _{0ik}\omega ^{k}=-E_{i}\omega ^{0}+0-\frac{1}{2}%
B_{k}\omega ^{j}+\frac{1}{2}B_{j}\omega ^{k},  \tag{83}
\end{equation}%
\begin{equation}
\beta _{ij}=\beta _{ij0}\omega ^{0}.  \tag{84}
\end{equation}%
In view of (84), the second line of (80) reduces to its first term, which
has to be exterior multiplied on the left by $\omega ^{i}.$ Hence, we first
compute $\beta _{0j}\wedge \beta _{0k}$ modulo $\omega ^{i}:$%
\begin{equation*}
\beta _{0j}\wedge \beta _{0k}=(-E_{j}\omega ^{0}-\frac{1}{2}B_{i}\omega
^{k})\wedge (-E_{k}\omega ^{0}+\frac{1}{2}B_{i}\omega ^{j})=\text{ \ \ \ \ \
\ \ \ \ \ \ \ \ \ \ \ \ \ \ \ \ \ \ \ \ \ \ \ \ \ \ \ \ \ \ \ }
\end{equation*}%
\begin{equation}
=-\frac{1}{2}E_{j}B_{i}\omega ^{0}\wedge \omega ^{j}+\frac{1}{2}%
E_{k}B_{i}\omega ^{k}\wedge \omega ^{0}+\frac{1}{4}B_{i}^{2}\omega
^{j}\wedge \omega ^{k},\text{ \ \ \ \ \ \ \ \ \ \ \ \ \ mod }\omega ^{i} 
\tag{85}
\end{equation}

On account of (77) and (80), we perform the left exterior multiplication of
(85) by $-\omega ^{i}$, set the factors in cyclic order and write out the
cyclic permutations:%
\begin{equation}
-\frac{1}{4}B^{2}\omega ^{ijk}+\frac{1}{2}(E_{j}B_{k}-E_{k}B_{jk})\omega
^{0jk}+\frac{1}{2}(E_{k}B_{i}-E_{i}B_{k})\omega ^{0ki}+\frac{1}{2}%
(E_{i}B_{j}-E_{j}B_{i})\omega ^{0ij},  \tag{86}
\end{equation}%
or, in compact form after dividing by $2\pi $,%
\begin{equation}
-\frac{B^{2}}{8\pi }\omega ^{123}+\frac{1}{4\pi }(E_{i}B_{j}-E_{j}B_{i})%
\omega ^{0ij},  \tag{87}
\end{equation}%
or%
\begin{equation}
\frac{B^{2}}{8\pi }\omega _{123}+\frac{1}{4\pi }(E_{i}B_{j}-E_{j}B_{i})%
\omega _{0ij},  \tag{88}
\end{equation}%
with summation over cyclic permutations. For comparison purposes with
standard theory, we also obtain the Hodge dual of these expressions, which
is a quantity of grade one, like the vector energy-momentum. We do this by
Clifford-multiplying (87) by $\omega _{0}\wedge \omega _{1}\wedge \omega
_{2}\wedge \omega _{3}.$ The result is%
\begin{equation}
\frac{B^{2}}{8\pi }\omega _{0}+\frac{1}{4\pi }(E_{i}B_{j}-E_{j}B_{i})\omega
_{k}.  \tag{89}
\end{equation}

For comparison with corresponding terms in classical electrodynamics, we
recall the equation 
\begin{equation}
\frac{\partial }{\partial t}\left( \dfrac{E^{2}+B^{2}}{8\pi }\right) =-%
\boldsymbol{j}\cdot \boldsymbol{E}-\mbox{div}\,\boldsymbol{S},  \tag{90}
\end{equation}%
where 
\begin{equation}
\boldsymbol{S}=\frac{1}{4\pi }\boldsymbol{E}\times \boldsymbol{B}  \tag{91}
\end{equation}%
is the Poynting vector. Whereas the absence of an $E^{2}$ term breaks the
symmetry in $E$ and $B$ in (87), the term $-\boldsymbol{j}\cdot %
\boldsymbol{E}$ breaks the same symmetry in (90). We are interested in the
pure-field terms, i.e. not containing the current, which should be expressed
in terms of the fields, more precisely in terms of general torsion. Then, in
view of view of (90), we should compare (88) with%
\begin{equation}
\frac{E^{2}+B^{2}}{8\pi }dV+\boldsymbol{S}\cdot \boldsymbol{f},  \tag{92}
\end{equation}%
where $\boldsymbol{f}$ indicates surface integration. We see that the term
in $E^{2}$ is missing. Where some readers may see this as a no-go result, we
see a call for the further development of differential geometry, along the
lines stated at the end of the previous section.

\section{Concluding remarks}

In this paper, we have shown the viability of TP for theoretical physics,
which Einstein tried to achieve in the late 1920's and failed. In so doing,
we have encountered another Einstein theme. In 1933, he proposed the thesis
of ---to use his own words--- logical homogeneity of theoretical physics and
differential geometry \cite{R38}.

In view of the remarks made about the need to generalize differential
geometry, this term must be understood here more comprehensively, i.e. based
not on the exterior calculus but on the K\"{a}hler calculus, the one with
which he developed an improved version of Dirac's relativistic quantum
mechanics. This will bring classical mechanics closer to quantum physics. We
shall also find out in our next paper through the use of this calculus for
the geometrization of the second pair of Maxwell's equations that quantum
mechanical issues such as the acquiring of mass by massless fields (Higgs
mechanism) already are a classical one. In the same paper, we shall also
undertake the development of the electromagnetic energy-momentum tensor and
a further discussion of the cosmological term. But it is only in K\"{a}%
hler's quantum mechanics where we have the mechanism (primitive constant
idempotents and the ideals they define) with which to construct particles
and thus matter \cite{R47}. 

In 1960-1962, K\"{a}hler already showed with his calculus that the
Copenhagen probabilistic interpretation need not be a foundational tenet of
quantum mechanics but an emergent one. The difficulty caused by the papers
being in German can be partially remedied with help of clarifiactions that I
posted in Alterman 2016, 2017 and 2018. The issue then is how can we use
this calculus to move the present paradigm in a direction which will
approach us towards something that we could call a theory of everything, or
at least a replacement more satisfying than what we have at present,
regardless of the latter's merits.

\end{document}